\documentclass[a4paper]{aa}  
\usepackage{graphicx}
\usepackage{amsmath}
\usepackage{amssymb}
\usepackage{natbib}

\usepackage{color}

\newcommand{\blue}[1]{{ \color{blue} #1 }}

\usepackage{txfonts}

\begin{document}

\title{False periods in complex chaotic systems}

\author{V. Votruba\inst{1,}\inst{2},
	P. Koubsk\'{y}\inst{1},
	D. Kor\v{c}\'{a}kov\'{a}\inst{1},
	F. Hroch\inst{2}}
\authorrunning{V. Votruba et al.}

\offprints{V. Votruba}

\institute{Astronomick\'{y} \'{u}stav, Akademie v\v{e}d \v{C}esk\'{e}
	republiky v.v.i., CZ-251 65 Ond\v{r}ejov, Czech Republic \\
	\email{votruba@sunstel.asu.cas.cz}
	\and 
	\'Ustav teoretick\'e fyziky a astrofyziky,
	P\v{r}irodov\v{e}ck\'{a} fakulta, Masarykova univerzita,
	Kotl\'{a}\v{r}sk\'{a} 2, CZ-611 37 Brno, Czech Republic}

\date{Received}

\abstract
{}
{Astrophysical objects frequently exhibit some irregularities or complex behaviour in their light curves. We focus primarily on hot stars, 
where both radial and non-radial pulsations are observed. One of the primary research goals is to determine physical parameters of stellar pulsations by analyzing their light curves or spectra, focusing on periodic or quasiperiodic behaviour.}
{We analyse the feasibility of classical methods for period searches in a nonlinear
chaotic system, such as the R\"{o}ssler system, where a
period does not exist at all.  As an astrophysical application of the chaotic
system, we utilize a simple model of stellar pulsation with two different sets of parameters corresponding to periodic and chaotic behaviour.  For both models we create a synthetic signal,
and then apply widely used methods for period finding, such as the
phase dispersion method and periodograms. For comparison, a
quasi-periodic signal is employed as well.}
{The period analysis indicates periods even for the chaotic signal. Such periods
are apparently spurious.  This implies that it is very problematic to
distinguish chaotic and quasiperiodic process by such an analysis only.
}
{}
\maketitle

\section{Introduction}
Variability of astronomical sources like stars is an important
aspect of their observation. We can detect variability in their 
light and/or spectra on different timescales. Dynamical pulsations, radial or 
non-radial, may be one of the main sources of intrinsic variability in many types of stars. As
was shown  by \cite{eddington}, stars may exhibit pulsation due to the
unstable layers in the stellar envelope. The first theoretical modelling of
pulsations predicted regular oscillations and such modelling was
succesfully applied to objects like classical Cepheids. But later, observations
indicated other types of pulsating stars, e.g. Population II Cepheids, which in
some cases exhibit irregular patterns in observational data. 

There was no reliable explanation of  such variability until the pioneering
work of \cite{baker}, which showed that a simple one-zone
model of stellar oscillators with a nonlinear adiabatic term can in some
specific conditions produce irregular pulsation. Similar results were also
obtained by \cite{moore}, \cite{auvergne1} and \cite{auvergne2}. The disadvantage of these 
models is their dynamically unstable equilibrium state, which seems to be
nonrealistic for stars. \cite{takeuti} and later \cite{saitou} showed that
similar, yet dynamically stable and more
precise one-zone model also produced chaotic solutions, with a
strange attractor similar to the R\"{o}ssler one \citep[see][]{rossler}. Detailed,
more realistic calculations of the state equation for helium ionization in
a one zone model, which was performed by \cite{buchler}, confirmed  that
pulsation instability may lead to chaotic behaviour. 

\cite{goupil} studied light curves of the pulsating white dwarf PG 1351+489 and 
found strong evidence of period doubling. Period doubling is a
scenario where a system that primarily oscillates with a fundamental frequency
$\nu_0$, after change of the control parameter oscillates with frequency ${\nu_0}/2$
\citep[see][]{helleman,gurel}. Cascades of period doubling is a
typical transition routes to chaotic behaviour. Such an effect is typical of chaotic systems. 
\cite{serre} analysed long time-series of photometric observations of the variable star R~Scuti, one
possible candidate pulsating star which  exhibits a strange irregular light
curve. By using the method of global flow reconstruction (see more about global
time series predictions in \cite{B1993}, \cite{abarbanel94}), they were able to
reproduce the observation with the chaotic dynamics very well. 
More details and examples of chaotic pulsations can be found in
\cite{buchlered}.
\begin{figure*}[ht!]
\scalebox{1.0}{\includegraphics{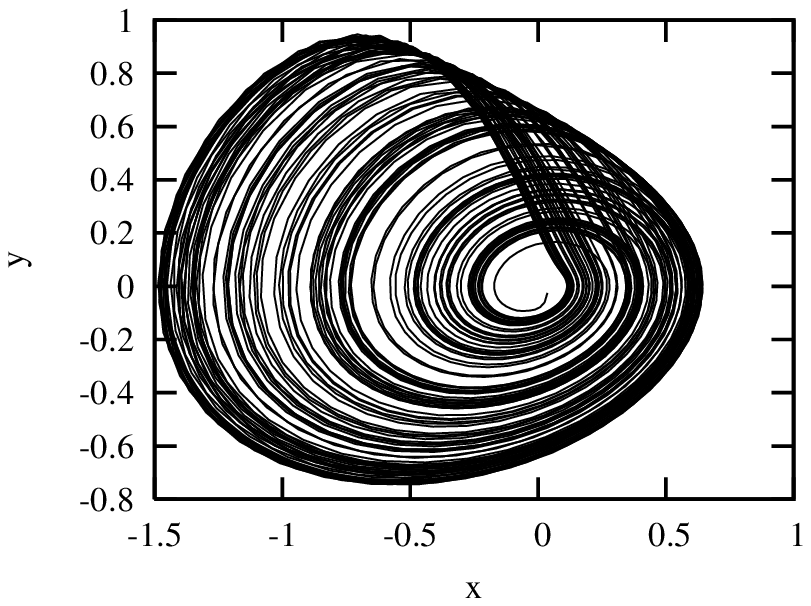}}
\scalebox{1.0}{\includegraphics{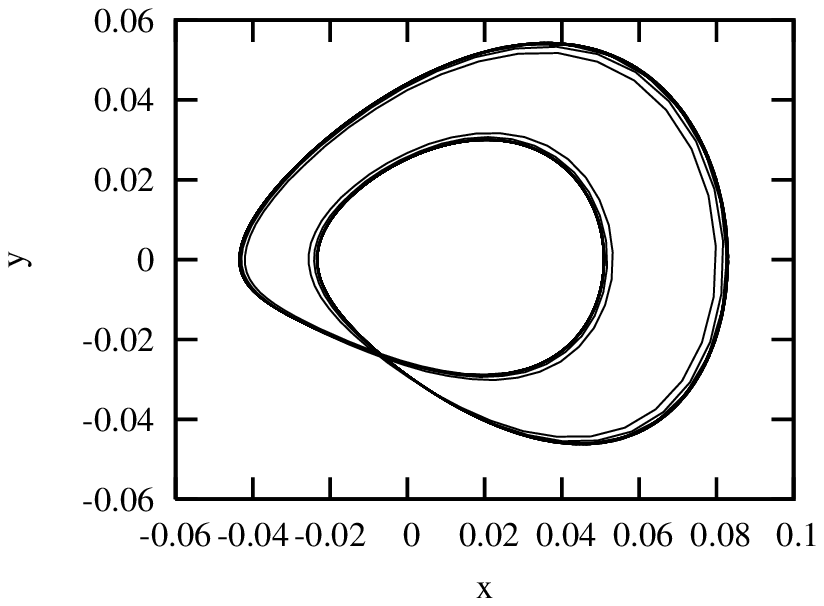}} 
\caption{{\it Left panel}: Chaotic attractor constructed from a numerical solution of the ODE system \eqref{ode} with the parameters from Table \ref{num}. - chaotic case. {\it Right panel}: Two-limit cycle attractor constructed from numerical solution of \eqref{ode} with the parameters from Table \ref{num}.  - periodic case. }
\end{figure*}
These clearly show that there is both theoretical and experimental evidence of quasiperiodic or chaotic behaviour in variable stars.  The problem is how to  distinguish between these two possibilities, if only limited time series are available,  often very unevenly sampled. In other words, what will be the result, if classical astronomical tools for the period searches, such as periodograms, the phase dispersion minimalization method or more sophisticated techniques are applied to data originating from quasiperiodic or chaotic sources. In the next paragraphs we tackle the problem using synthetic data.
\section{Generation of synthetic time series}
As an example of irregular stellar variability due to pulsation we use
 the model of \cite{tanaka}. In this model, the set of
ordinary differential equations (ODE) for a stellar oscillator with nonlinear, nonadiabatic terms, reads
\begin{eqnarray}
\nonumber \frac{{\rm d}x}{{\rm d}t} &=& y \\
\nonumber \frac{{\rm d}y}{{\rm d}t} &=& \alpha + \mu y + z \\
\frac{{\rm d}z}{{\rm d}t} &=& -\beta y - p z - q y + s y z 
\label{ode}
\end{eqnarray}
where  the first two equations represent equations of motion and the last one is a state equation with a nonadiabatic term.  We can adjust the behaviour of the model  using different values of $\alpha,\beta,\mu,p,q,s$. The system of ODE {of the first order} was numerically solved with the Runge-Kutta method for two different sets of model parameters,  the first one corresponding to the periodic regime with a two-limit cycle attractor and the  second one  to the aperiodic, chaotic regime with a strange attractor similar to the R\"{o}ssler type. The parameters are summarized  in Table \ref{num}. \begin{figure*}[t!]
\scalebox{1.0}{\includegraphics{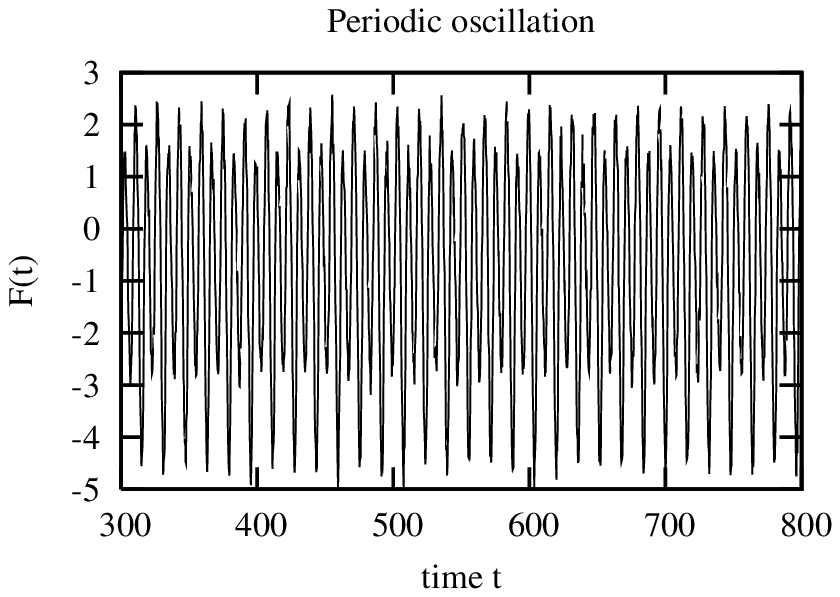}}
\scalebox{1.0}{\includegraphics{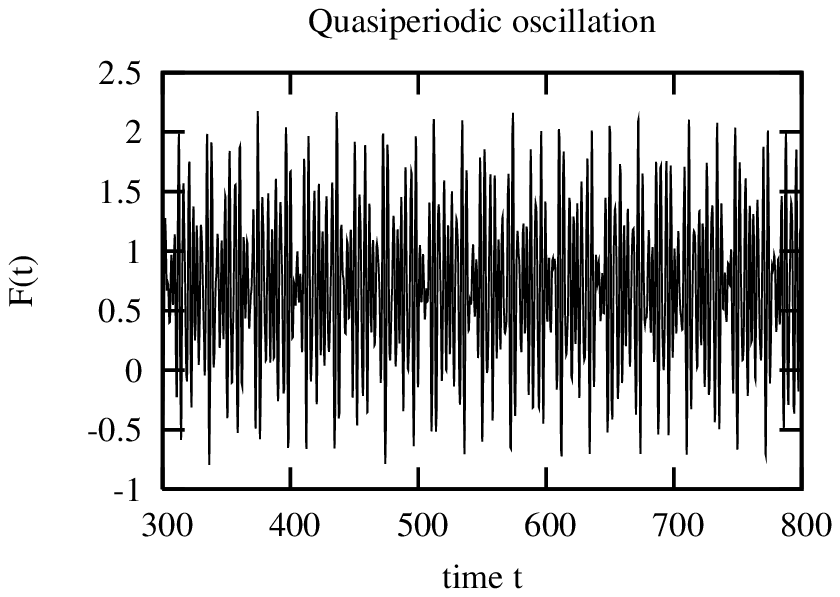}} \\
\scalebox{1.0}{\includegraphics{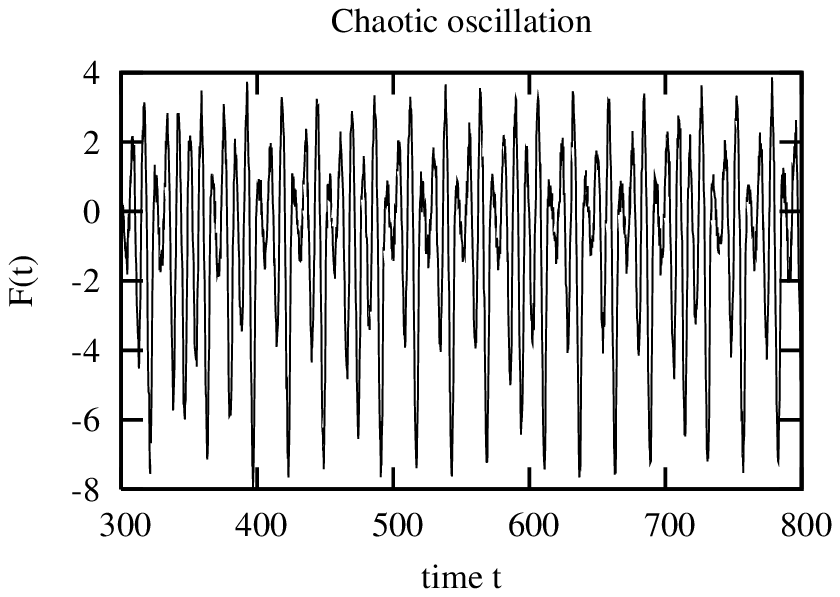}}
\scalebox{1.0}{\includegraphics{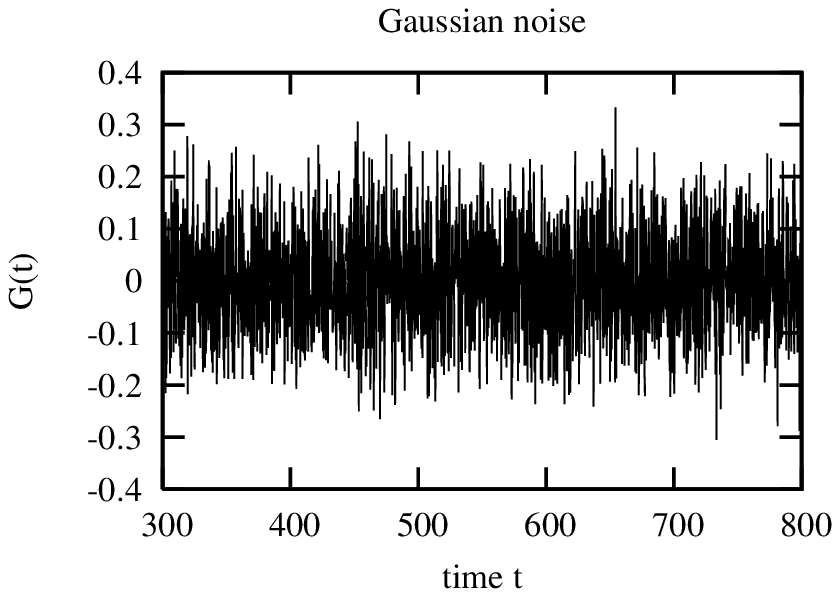}}
\label{gsolution}
\caption{{\it Upper left panel:} $x(t)$ solution of the ODE system \eqref{ode} for the periodic case with a two-limit cycle. {\it Upper right panel:} $x(t)$ solution of equation \eqref{quasi} for the quasiperiodic case. {\it Lower left panel:} $x(t)$ solution of the ODE system \eqref{ode} for the chaotic case.  {\it Lower right panel:}  Gaussian noise (\ref{gaussian}) with parameters  ${\sigma}=0.1$ and $\overline{x}=0$. All physical quantities are in arbitrary units.}
\end{figure*}
We use the initial conditions ($x_0 = 0.37$, $y = -0.249 $, $z = -0.174 $) for the chaotic solution.
The initial conditions are very close to the orbit of the
attractor in phase space. We obtain a time-dependence of the state variables $x(t),y(t),z(t)$ in the form of
 a time series with an equidistant time step. We choose one state variable,
namely $x(t)$, as a representation of the measured variable. To simulate a synthetic signal from the quasiperiodic process, a different approach is employed. 
The signal is modeled by the following
equation
\begin{equation}
x_{\rm quasi}(t) = a_1\sin{(f_1 t)}+a_2\sin{(f_2 t)}+a_3\sin{(f_3 t)}
\label{quasi}
\end{equation}
where $a_1,a_2,a_3$ are amplitudes of incommensurate modes with frequencies
$f_1,f_2,f_3$. Their values are  summarized  in Tables \ref{num} and
\ref{quasnum}.  These parameters are carefully chosen in order to obtain  
similar dependencies of $x(t)$ and $x_{\rm
quasi}(t)$ for the chaotic and quasiperiodic case, namely amplitudes of
oscillations (see Fig. \ref{gsolution}). The computed signal for each type of
process is plotted in Fig. \ref{gsolution}. Since we want to simulate
real measurements, we add a synthetic signal with Gaussian noise. It is
important that for a simulation of light curves from objects others
than stars, such as QSOs, noise may by colored and then it is necessary to use a
different type of noise, e.g. Poisson noise. The noise can be described using a
normal distribution
\begin{equation}
G(x)=\frac{1}{\sigma \sqrt{2\pi}}\exp{((x-\overline{x})^2/{2{\sigma}^2})}
\label{gaussian}
\end{equation}
where $\sigma$ is the variance and $\overline{x}$ is the mean value of the
measurement errors.  For numerical simulations of the noise with a normal
distribution, we use the method of \cite{NumRand}. It was necessary to set the different values of $\sigma$ for
each type of signal in order to have similar S/N ratios.
\begin{table}[h!]
\caption{Parameters of the stellar pulsation model}
\begin{tabular}{c c c c c c c c c}
\hline
 & $\alpha$ & $\beta$ & $\mu$ & p & q & s & $\sigma$ & $\overline{x}$ \\ 
\hline
periodic & -0.5 & 0.5 & 0.5 & 3.2 & 0.5 & 1.0 & 0.2 & 0\\
chaotic & - 0.5 & 0.5 & 0.5 & 4.0 & 0.5 & 1.0 & 0.3 & 0\\
\hline
\end{tabular}
\label{num}
\end{table}
Data for  synthetic time series satisfy the condition $t>300$, i.e.
reasonably far away from the starting point ($t=0$). It is necessary to make
 sure that the solution is relaxed away from the influence
of the initial conditions. 
 
 It can be readily seen that chaotic and quasiperiodic oscillations (see Fig. \ref{gsolution}) have very similar patterns, different to the simple periodic signal (two-limit cycle).

\begin{table}[b]
\caption{Parameters for the quasiperiodic signal}
\begin{tabular}{c c c c c c c c }
\hline
$a_1$ & $a_2$ & $a_3$ & $f_1$ & $f_2$ & $f_3$ & $\sigma$ & $\overline{x}$ \\ 
\hline
0.4 &0.6 & 0.5 & $\sqrt{5}$ & $\sqrt{3}$ & $\sqrt{2}$ & 0.1 & 0\\
\hline
\end{tabular}
\label{quasnum}
\end{table}
\section{Analysis of the signal}
\subsection{Classical period analysis}
The most common techniques used for period searches are the power spectral
analysis (PWS), the modified Lomb periodogram (LPD), the phase dispersion
minimalization (PDM), and the string length method (SLM). In
the first case, discrete Fourier transformation is applied to a dataset
resulting in a power spectrum \citep[for more details see e.g.][]{scargle}.
For unevenly sampled series it is better to use a modification of PWS,  method
LPD \citep[see for more details][]{lomb}. The third method
binds data in a grid of phases and minimizes phase
dispersion statistics \citep[see for more details][]{stellingwer,lafler}. The SLM
method is similar to the previous one \citep[see][]{dworetsky}, and the
LPD method behaves like the PWS. So we limit on the two methods, PWS and PDM
only. Obviously,  PDM and PWS techniques applied on a periodic signal
detect corresponding frequencies without any doubt, but,  as we will show,
 a different situation arises for quasiperiodic and chaotic
regimes.

\begin{figure*}[h!]
\scalebox{1.0}{\includegraphics{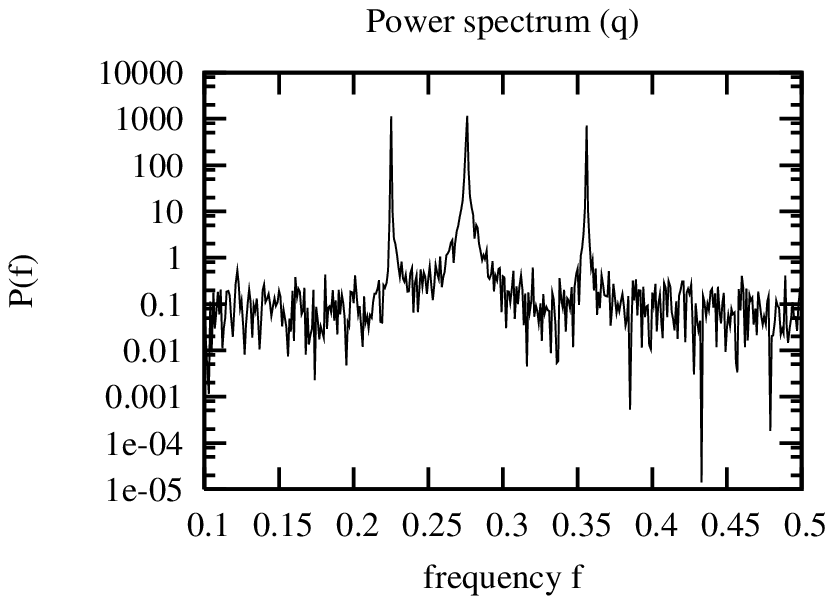}}
\scalebox{1.0}{\includegraphics{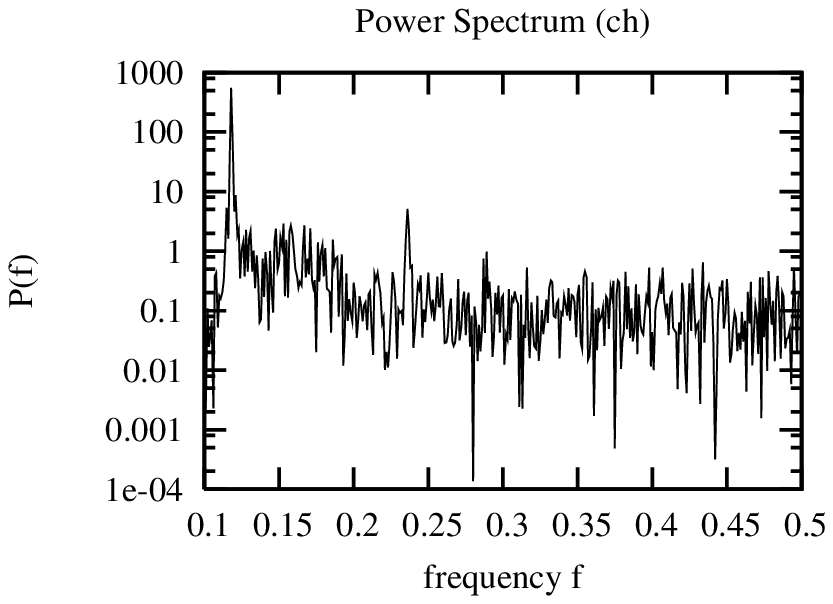}} \\
\scalebox{1.0}{\includegraphics{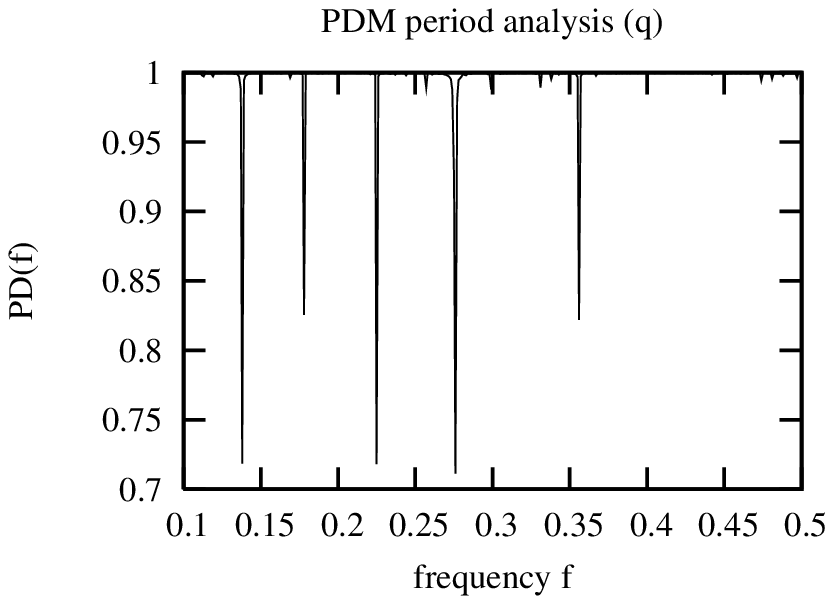}}
\scalebox{1.0}{\includegraphics{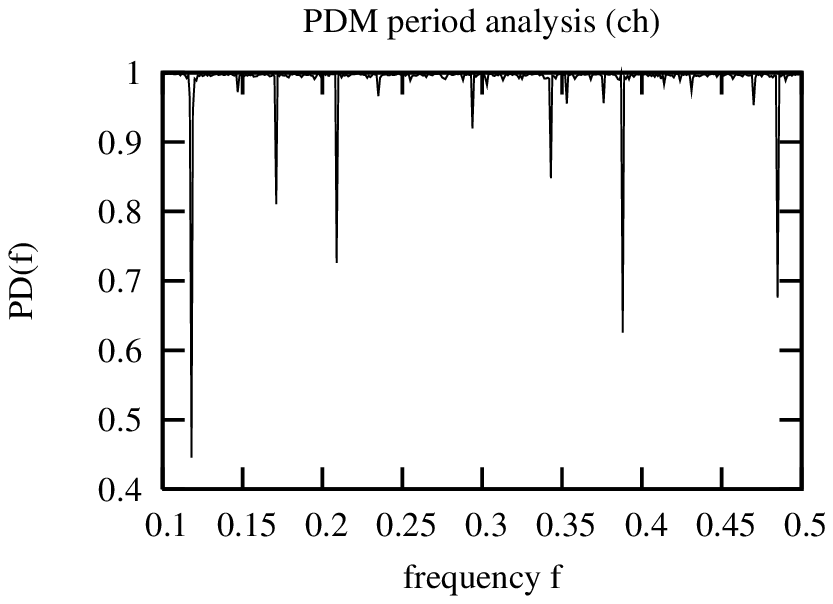}}
\caption{{\it Upper panels}: Power spectra constructed for quasiperiodic oscillation ({\it on the left}) described by \eqref{quasi} and the chaotic case of stellar oscillation ({\it on the right}) described by \eqref{ode}. {\it Bottom panels}: Periodogram constructed using the phase dispersion minimalization method for the quasiperiodic case ({\it on the left})  and chaotic case ({\it on the right}). For the frequency we used arbitrary units.}
\label{powerpdm}
\end{figure*}

\begin{figure*}
\scalebox{1.0}{\includegraphics{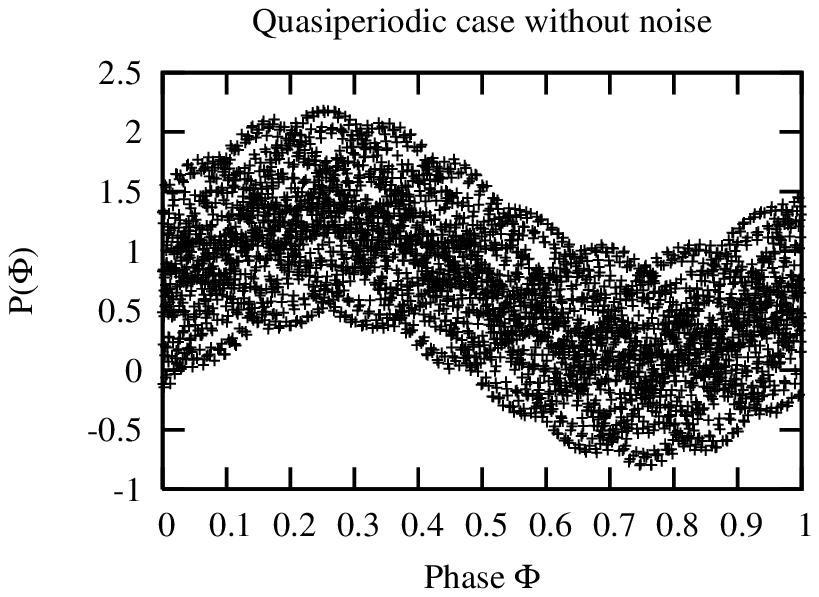}}
\scalebox{1.0}{\includegraphics{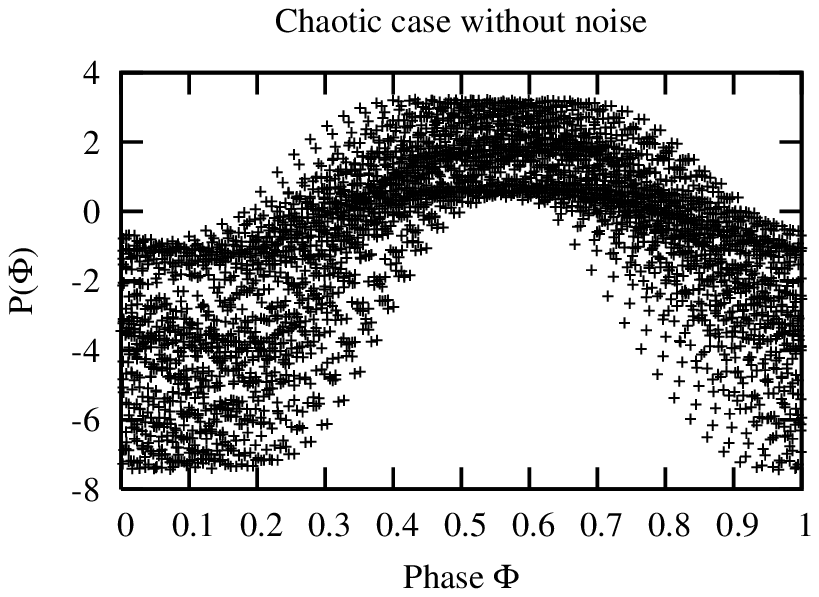}}  \\
\scalebox{1.0}{\includegraphics{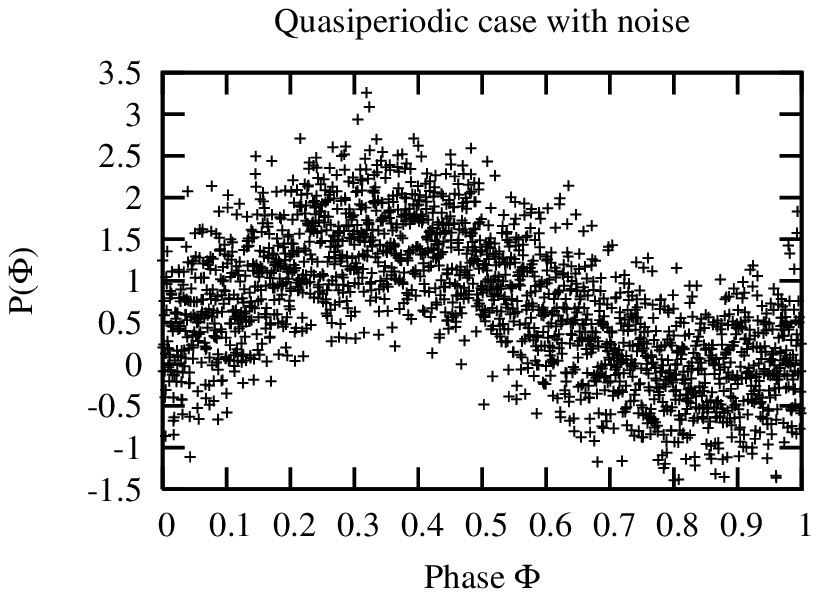}}
\scalebox{1.0}{\includegraphics{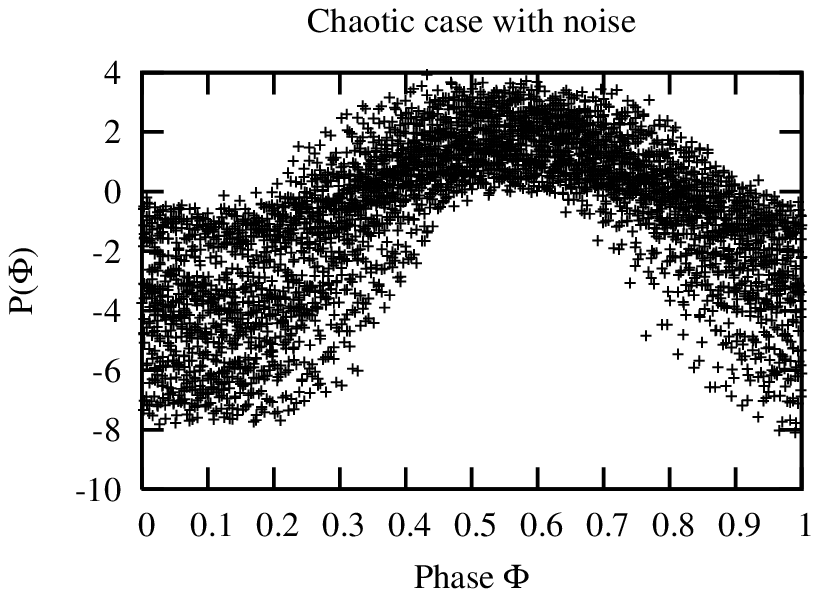}} 
\caption{{\it Left upper panel:} Phase diagram for the quasiperiodic case and the dominant frequency $f_1=\sqrt{3}$ without noise.
{\it Right upper panel:} Final phase diagram for the frequency $f_1=0.1176$ of the chaotic case. 
{\it Left bottom panel}: The same as the left upper panel but with Gaussian noise. {\it Right bottom panel:} The same as the right upper panel,
but with Gaussian noise. On the bottom panels both diagrams are similar, showing a simple sine-wave pattern. All physical quantities are in arbitrary units.}
\label{phasedia}
\end{figure*}

{\it 1. Quasiperiodic signal.} The PWS analysis of the quasiperiodic signal
produced by known incommensurate frequencies (see Table \ref{quasnum})
detects the frequencies used.
Moreover, it is possible to recognize the leading
frequency, which is in our case $f_2=\sqrt{3}$. The noise
in the periodogram in Fig. \ref{powerpdm} is caused by the simulated Gaussian
noise of the signal.  The PDM method yields a similar
estimate of the input frequencies, and the results also show some other
frequencies that are subharmonics or harmonics of the frequency set. The phase
diagram in Fig. \ref{phasedia} (on the left) is folded with the leading
frequency $f_2$. Clearly, the structure is not a simple sine wave.
It is a quasiperiodic, and hence it is composed of a set of harmonics with slightly different phase. 
After accounting for Gaussian noise, this difference is smeared out.

{\it 2. Chaotic signal.}
The results become much more interesting for the chaotic signal.
Again, both PWS and PDM  methods detect periods, as can be clearly seen in Fig. \ref{powerpdm}. After some checks 
we can recognize the two longest independent periods.The power spectrum of the chaotic signal shows the
most significant frequencies together with the frequency noise.
Nevertheless, low-level noise is present even if Gaussian
noise is not applied. It is the consequence of the chaotic behaviour.
{
}
\begin{figure*}[t!]
\scalebox{1.0}{\includegraphics{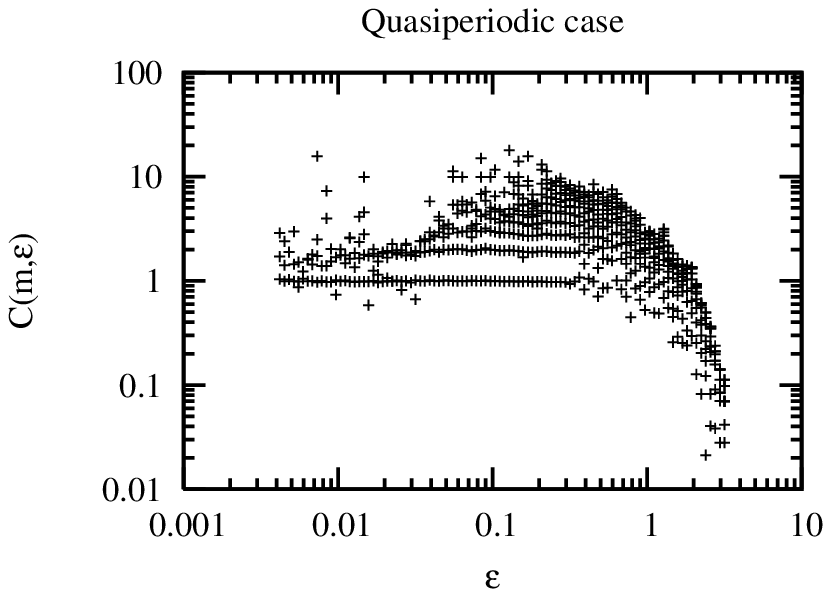}}
\scalebox{1.0}{\includegraphics{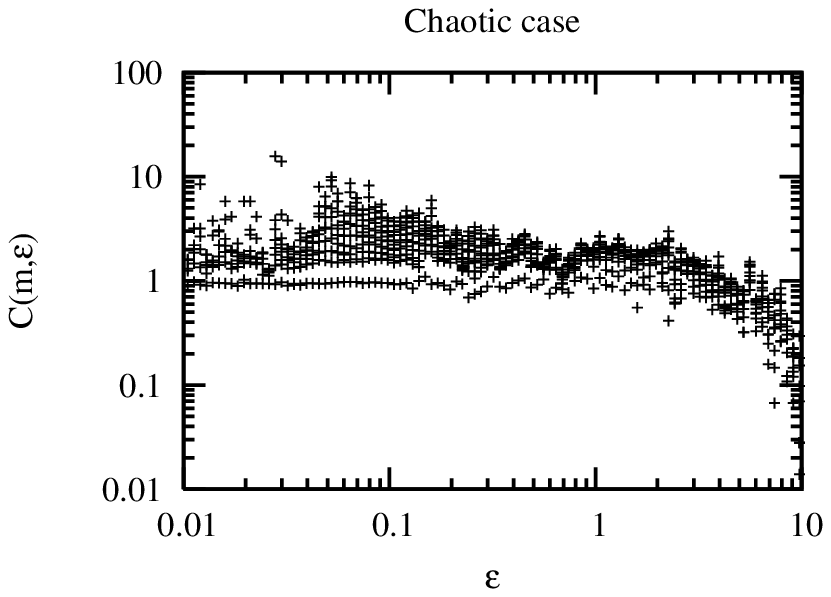}} 
\caption{correlation integrals {\eqref{cd}} for different lengthscales $\epsilon$ and embedding dimensions $m$. The quasiperiodic case is in the left panel, chaotic case is in the right panel.}
\label{cdimension}
\end{figure*}

Theoretically, power spectra of the chaotic data are composed of an infinite number
of frequencies. This is true, however,  only if there is an infinite number of
observations \citep[see][chap. 5]{regev}. In reality, there is  only a limited number of
observations,which are, furthermore, unevenly spaced. The phase diagram constructed for the
 most significant frequency derived $f_1=0.1179$ is shown in Fig.
{\ref{phasedia}} (on the right). In the phase diagram without Gaussian noise, we
can recognize a two-wave structure in the phase plot. On the other hand, when
Gaussian noise  is applied a simple noisy sine wave can be identified. This is
similar to the quasiperiodic case results, although the data originated
from a nonlinear chaotic process, where period has no meaning.  Other
frequencies $f_2,f_3$ lead to even less clear phase diagrams.

{The above results clearly show that it is very difficult to distinguish between chaotic and quasiperiodic cases using period analysis only. Thus, we must look for other methods, which can help to resolve both processes.}

\subsection{Nonlinear analysis}
{For detailed  analysis of the signal that can be generated by nonlinear
chaotic processes, we can use methods of nonlinear time series analysis. These
methods were developed for classification and identification of chaotic
phenomenon in time series of the experimental
data \citep[see review][]{abarbanel}.} As a indication of chaotic behaviour we can use topological
invariants, namely the correlation dimension and Lyapunov exponents. Algorithms for
numerical estimates of such invariants from time series are described by
\cite{abarbanel} and by \cite{tisean}. In our case the indicator of chaotic
behaviour is the correlation integral defined (see more detail \cite{tisean})
as
\begin{equation}
C(m,\epsilon)=\frac{2}{(N-n_{\rm min})(N-n_{\rm min}-1)}\sum_{i=1}^{N}\sum_{j=i+1+n_{\rm min}}^{N}\theta(\epsilon-||{\bf s}_i-{\bf s}_j||)
\label{cd}
\end{equation}
where $m$ is the  embedding dimension, $\bf{s}_n$ denotes the delay embedding
vector, $\epsilon$ is the lengthscale parameter{\blue{,}} and $N$ is the number of
points. A possible problem with temporal correlations is overcome by excluding
pairs that are too close in time (parameter $n_{\rm min}$). Before we apply
the procedure for calculation of the correlation dimension to the noisy signal
(chaotic or quasiperiodic), it is necessary to clean the noisy data. This is done by
 using nonlinear noise reduction with locally constant
approximations - the simple nonlinear filter described by \cite{schreiber}. The
results of the calculation are shown in Fig. \ref{cdimension}. The chaotic
processes are characterized  by a plateau in the  plot of $C(m,\epsilon)$ for
different embedding dimensions (left panel Fig. \ref{cdimension}), which
indicates a self-similar geometry. A linear fit to this plateau region leads to
the estimation of correlation dimension
\begin{equation}
C_{\dim}=1.90 \pm 0.08.
\end{equation}
This value agrees with the typical value of the R\"{o}ssler type attractor. In the case of the quasiperiodic signal we cannot identify the plateau region (right  panel Fig.\ref{cdimension}). This indicates that there is no self-similar geometry for this type of the signal.

\section{Summary}
Analysis of synthetic signals simulating quasiperiodic and chaotic data led to the following findings:

1) In the realistic case ( finite number of observations with Gaussian noise), it is impossible to distinguish between quasiperiodic and chaotic behaviour by classical methods of period analysis. 

2) When the data contain large amounts of noise, the chaotic or quasiperiodic character of the signal can be overlooked and  a simple periodic solution with the strongest frequency can be determined from the period analysis.

3) In order to definitely rule out presence of chaos it is neccessary to use a nonlinear time series analysis. Thus, we must reconstruct the topologically similar phase portrait of the system using the technique described by \cite{abarbanel}. Then we must determine the invariants, such as the correlation dimension or Lyapunov coefficients from the reconstructed time series, and from their values we can decide if deterministic chaos is present in the data or not. However, both these procedures need better quality of data (noise, length of time series).

4) We would like to draw attention to the fact that irregular chaotic behaviour may be more usual in astronomical data than  thought before, and that some systems that were identified as quasiperiodic may actually be chaotic ones.
\begin{acknowledgements}
We thank S. Shore and T. Dytrych for helpful comments on our analysis. This research has made use of NASA's Astrophysics Data System.
This work was supported by the grant CZ 205/09/P476. The Astronomical Institute Ond\v{r}ejov is supported by project AV0\,Z10030501.
\end{acknowledgements}

\end{document}